\documentclass[10pt, conference, final]{IEEEtran}

\usepackage{cite}
\usepackage{amsmath,amssymb,amsfonts}
\usepackage{amsthm}
\usepackage{wrapfig}
\usepackage{algorithm}
\usepackage{algpseudocode}
\usepackage{graphicx}
\usepackage{textcomp}
\usepackage{xcolor}
\usepackage{subcaption}
\usepackage{multirow}
\usepackage{booktabs}
\usepackage{soul}
\usepackage{array}
\usepackage{float}
\usepackage{tikz}
\usepackage{longtable}

\begin{document}
\title{Zero Day Threat Detection Using Metric Learning Autoencoders}

\author{Dhruv Nandakumar$^{a}$,
        Robert Schiller$^{a}$,
        Christopher Redino$^{*}$$^{a}$,\\
        Kevin Choi$^{a}$,
        Abdul Rahman$^{a}$,
        Edward Bowen$^{a}$,
        Marc Vucovich$^{a}$\\
        Joe Nehila$^{a}$,
        Matthew Weeks,
        Aaron Shaha\\
        \small $^{a}$Deloitte \& Touche LLP \\
        \small $^{*}$Corresponding author: credino@deloitte.com \\
}

\maketitle

\begin{abstract}

The proliferation of zero-day threats (ZDTs) to companies' networks has been immensely costly and requires novel methods to scan traffic for malicious behavior at massive scale. The diverse nature of normal behavior along with the huge landscape of attack types makes deep learning methods an attractive option for their ability to capture highly-nonlinear behavior patterns. In this paper, the authors demonstrate an improvement upon a previously introduced methodology, which used a dual-autoencoder approach to identify ZDTs in network flow telemetry. In addition to the previously-introduced asset-level graph features, which help abstractly represent the role of a host in its network, this new model uses metric learning to train the second autoencoder on labeled attack data. This not only produces stronger performance, but it has the added advantage of improving the interpretability of the model by allowing for multiclass classification in the latent space. This can potentially save human threat hunters time when they investigate predicted ZDTs by showing them which known attack classes were nearby in the latent space. The models presented here are also trained and evaluated with two more datasets, and continue to show promising results even when generalizing to new network topologies.

\end{abstract}

\section{Introduction}
Cyber attacks on enterprise networks are dramatically increasing in frequency, scale, and complexity \cite{dunsavage_2022}. The year 2021 saw a record breaking 623 million cyber attacks globally, a 98\% increase over the previous year \cite{edwardson_2022}. A large proportion of such attacks are performed using novel tactics, techniques, and procedures (TTPs) and are known as Zero-Day attacks or threats. Henceforth, we define ZDTs as any novel TTP used with malicious intent on cyber systems. While current trends in cyber-security indicate that organizations are steadily detecting more ZDTs every year, the types of vulnerabilities discovered as a cause of said attacks remained the same \cite{stone_1970}. Traditional signature-based approaches to identifying cyber attacks are, by definition, designed to detect known patterns of threats and are ineffective against completely novel threats. Furthermore, the immense volume and intricacy of cyber-security telemetry make manual, unaided, threat hunting infeasible for security operators. These factors necessitate the use of more advanced anomaly detection, such as those offered by modern machine learning techniques, that can quickly and reliably analyze petabyte-scale data volumes for potential ZDTs. 

Machine learning based methods for cyber anomaly detection have shown promising results in ZDT detection by successfully baselining 'normal' network flow behavior in order to detect abnormal behavior that may be indicative of a cyber attack. However, while these approaches are effective at cyber anomaly detection, they tend to generalize to novel network topologies poorly and are also inherently ineffective at differentiating between known cyber anomalies and ZDTs. Furthermore, the increasing variety of cyber attack TTPs have also proven challenging for traditional anomaly detection approaches, which tend to perform poorly when identifying anomalies outside of classes with low support during training. In this paper, we introduce a novel technique for ZDT detection on network-flow telemetry that has demonstrably detected ZDTs on real enterprise-telemetry. The technique utilizes a dual-autoencoder architecture for anomaly and novelty detection respectively and has shown strong performance in identifying novel cyber threats in new network topologies without retraining. Furthermore, we augment the dual autoencoder approach with metric learning to help the models adapt to high attack type variety and class imbalance during inference. Overall, the key contributions of this paper are as follows:

\begin{enumerate}
    \item A dual-autoencoder based approach to ZDT detection that utilizes metric learning to segregate novel threats from those seen during training despite low support and high attack variety and achieves stronger performance than other ZDT detection approaches seen in literature.
    \item A novel approach to attack type attribution and model interpretability that enables security operators to identify the closest attack type to a ZDT for easy threat hunting.
    \item Models are trained and evaluated on enterprise scale security telemetry and have shown success in generalizing and identifying ZDTs in near-real time on novel network topologies. 
\end{enumerate}

The remainder of this paper includes a Literature Review discussing previous work in the ZDT detection field so
far, followed by a brief overview of the concepts covered in the work. We then describe the methodology's implementation details, experimental design, and results before closing with discussion and conclusion sections.

\begin{table*} [t!]
    \centering
    \caption{Datasets Used For Training and Evaluation}
    \label{table:1}
\begin{tabular}{|p{0.1\linewidth}|p{0.35\linewidth}|p{0.45\linewidth}|}
    \hline
    Dataset & Description & Notes \\
    \hline \hline
    MAWI Archive (MAWI) & Real network packet capture (PCAP) data from hundreds of devices across 12 universities in Japan. & Flow data for 7 consecutive days from 2021, and 1 day from 2016 were used from a dataset of over 14 years. Dataset is not labelled but is considered benign due to the extreme class imbalance of anomalies in real-world networks.  \\
    \hline
    National Collegiate Cyber Defense Competition (NCCDC) & Consists of labelled blue-team and red-team data from real attack simulations on a cyber range. & Data spanned two consecutive days in 2020 and in 2021. Attack types consist of network scanning, interrogation, botnet, and command and control. \\
    \hline
    ISOT Botnet Dataset (ISOT) & Data consists of benign and botent network traffic. & Botnet activity comes from the Storm and Waladec botnets. This is merged with captured real-world benign traffic from Lawrence Berkeley National Laboratory (LBNL) to produce a dataset with both benign and malicious activity \cite{isot_2010}. \\
    \hline
    ISCX IDS Evaluation Dataset (ISCX) & Data consists of benign and botnet network traffic. & Real-world traffic was analyzed to generate profiles for different types of agents. These profiles were then used to generate synthetic traffic mimicking the real agents and containing both benign and malicious activity \cite{iscx}. \\
    \hline
    Organizational Internal Flow (OIF) & Data consists of real internal network traffic from multiple locations. & Flow data for 7 consecutive days was used. Dataset is not labelled but is considered benign due to the extreme class imbalance of anomalies in real-world networks. \\
    \hline
    Organization Malware Lab (Codex) & Data consists of network data of over 100 million flows over 5 years from hundreds of real malware sample detonations in our internal malware cyber-range. & Dataset is labelled with malware class name, and was correlated with threat intelligence to extract higher level attack-type labels including botnets, ransomware, infostealer, e.g.. Data did not contain non-malicious flows.\\
    \hline \hline
\end{tabular}
\end{table*}

\section{Literature Review}
While statistical methods have long been used to catch known cyber threats, their utility for systematically identifying ZDTs has recently gained attention. There has been work in detecting ZDTs using methods including, but not limited to, rules-based approaches as well as unsupervised clustering methods using malware opcode sequences \cite{opcode} and Bayesian classifiers \cite{baysean}. Many authors have focused, as we do in this work, on using network flow data to detect ZDTs. Lobato, Lopez, et al. \cite{adaptiveZero}  utilized supervised machine learning approaches such as support vector machines to classify network flow telemetry as malicious or benign using an adaptive data modeling and pipeline approach with promising results, achieving precision and recall values between $0.66$ and $0.97$. Blaise et al.\cite{portbased} utilized network flow telemetry to identify ZDTs using an unsupervised approach that identifies anomalous port usage. Sarhan et al.\cite{zeroShot} also utilized network flow telemetry to identify ZDTs using a zero-shot learning approach with random forest and neural network models using a novel data splitting approach to hold out attack classes as ZDTs to measure performance.

Autoencoders have shown particular promise for identifying ZDTs because of their success as anomaly detectors. An autoencoder is a neural network consisting of an \textit{encoder}, whose layers successively compress the size of the input data, and a \textit{decoder}, which attempts to reconstruct the input from the compressed version. By training with only normal data, the autoencoder is taught to correctly reconstruct normal events and should produce a higher mean-squared error between the input and reconstruction when predicting anomalous data. Hindy et al.\cite{vanillaAutoEnc}, Yousefi-azar et al.\cite{yousefiAE}, and An and Cho\cite{VAE} demonstrated the effectiveness of autoencoders trained on network telemetry data for intrusion detection, achieving strong precision and recall values compared to traditional methods such as RF models. Zhang et al.\cite{semanticAE} also utilized a modified autoencoder with semantic segmentation of attack types in natural language to identify ZDTs with strong accuracy values up to $0.88$ as compared to other methods. However, the above studies are all limited to single autoencoder models performing anomaly detection after being trained on benign data, and only utilize flow-based features or semantic representations of said features. Redino et al. previously demonstrated the efficacy of a dual-autoencoder approach at identifying ZDTs, especially when paired with asset-level graph features which help abstract the roles of individual Internet Protocol (IP) addresses to multiple networks; this model was able to achieve precision and recall values of $0.83$ and $0.63$, respectively, on realistic testing data, outperforming both traditional methods and naive deep learning models \cite{redino_2022}.

\begin{figure*}[h]
    \includegraphics[width=\textwidth]{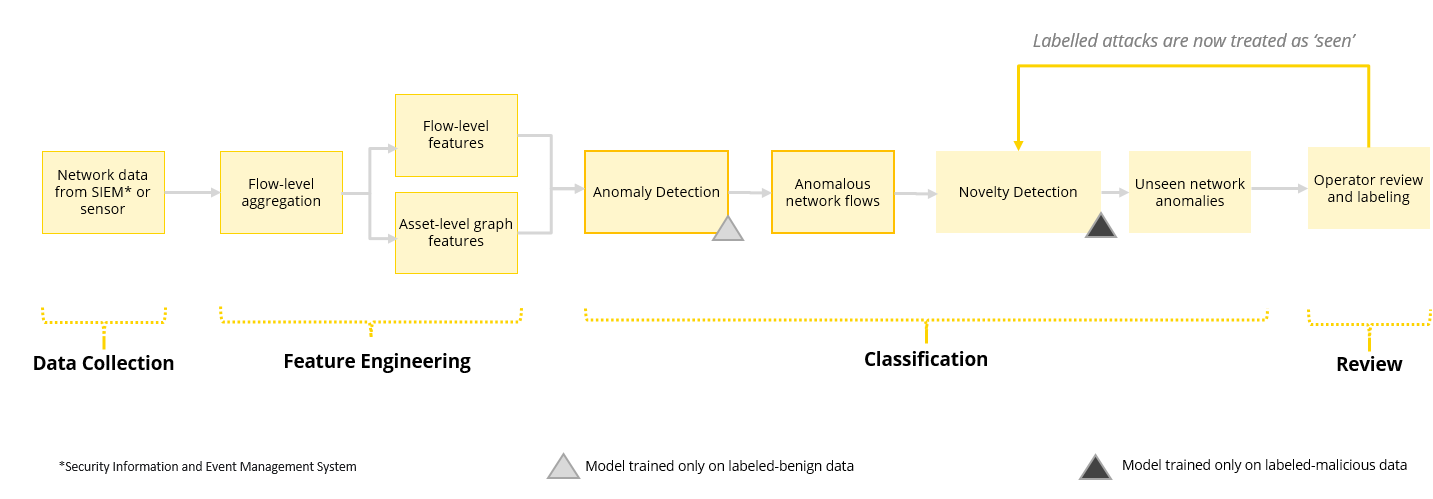}
    \centering
    \caption{High-Level Architecture \cite{redino_2022}}
    \label{fig:modelArch}
\end{figure*}

While autoencoders have generally performed well at identifying novel threats, they have no inherent concept of the different classes of threats they see in training. This means that the \textit{latent space}, the space containing the compressed data after passing through the encoder, can be chaotic, with vectors representing completely different classes of behavior placed near each other. Forcing the network to produce a more organized latent space, with clusters of similar behavior, presents several advantages. It teaches the network to extract the similarities between events within a particular class, and it helps to avoid posterior collapse, a situation where the encoder compresses the input data to an even smaller dimension than desired and destroys necessary information for the decoder. It also admits the possibility of performing multiclass classification on events based on their latent-space representations as a post-processing step, which could be useful to operators trying to understand the model's final output by providing added context of attack type attribution. Deep metric learning can be used to introduce order to the latent space by forcing models to make meaningful embeddings of classes by separating and clustering them, and has already gained significant traction for outlier detection in fields such as image recognition. In the cyber domain, Qu et al. specifically showed that a single autoencoder trained with metric learning could learn to identify anomalous behavior in Transmission Control Protocol (TCP) traffic better than baseline models \cite{qu_2020}. Andresini, Apprice, and Malerba used a combination of two autoencoders and a separate classifier trained with triplet learning to identify known network intrusions, outperforming various state-of-the-art Intrusion Detection Systems (IDS) models on the NSL-KDD dataset \cite{andresini2021}. Similarly, Wang et al. proposed a fully-connected encoder which simultaneously optimizes the multiclass classification loss and triplet loss, to achieve an accuracy score of 0.81 on the same dataset \cite{wang_2022}. Note that neither of these previous two approaches are seeking ZDTs, but their success further demonstrates the promise of metric learning for attack class discrimination.

A significant drawback to many of these approaches is the scarcity of realistic, labeled attack data for training and evaluation. The commonly-used academic datasets do not contain sufficient diversity of attack types to properly test how the model performs on ZDTs. We aimed to test our models on a variety of attack types and with data from multiple different networks to ensure robustness and generalizability to new data.

\begin{figure*}[h]
    \includegraphics[width=\textwidth]{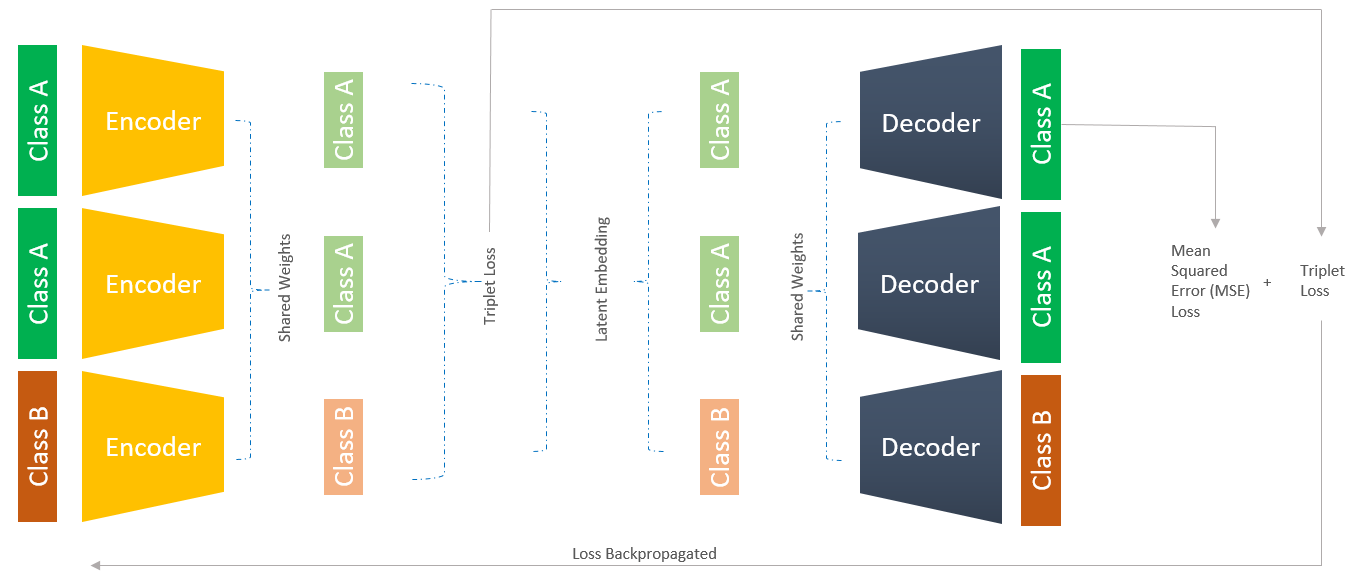}
    \centering
    \caption{Novelty Detector Triplet Training Architecture}
    \label{fig:tripletArch}
\end{figure*}

\section{Methodology}

\subsection{Datasets}
Datasets were required to contain flow-level information about each event (connection duration, port numbers, timestamp, and number of forward and backward bytes transmitted), the source and destination IP addresses, and attack class labels. IP addresses are necessary to generate the asset-level graph features, so some publicly-available datasets with labeled malicious traffic, such as CICIDS-2018, were excluded. Similarly, the attack labels must be sufficiently granular to perform metric learning, so datasets with binary labels were also excluded. In total, we used four publicly-available datasets as well as two proprietary datasets, as summarized in Table \ref{table:1}.

\subsection{Feature Engineering}
This works aims to augment the approach Redino et al. \cite{redino_2022} proposed with metric learning, and consequently utilizes the same feature engineering approach for ZDT detection. Our ZDT detector requires only connection-level information that can be obtained from network flow logs. Using the connection source and destination IP addresses, we construct an overall graph of connections, where each connection between hosts contributes edge weight 1. We then use this interaction graph to compute 9 graph features to describe each host, with the goal of abstracting the role that different types of hosts play in their communities. For each event, then, we append to the flow-level data these asset-level graph features for both the source and destination IP addresses, as well as a flag which detects when a connection crosses between communities in the graph. Finally, the actual IP addresses are not included as features to avoid overfitting to the training networks.

\subsection{ZDT Model Architecture}
Redino et al.'s \cite{redino_2022} dual-autoencoder approach splits the task of identifying ZDTs into \textit{anomaly detection} and \textit{novelty detection}:
\begin{enumerate}
    \item \textbf{Anomaly Detector (AD)}: Trained on benign network traffic and designed to identify any malicious activity that effect the network flow. All events pass through the AD at inference; events that exceed the loss threshold are considered anomalous.
    \item \textbf{Novelty Detector (ND)}: Trained on known attack types and designed to identify previously-unseen attack types. Only events that have already been labeled anomalous by the AD pass through the ND at inference, though in practice this data will still contain some benign traffic. The key addition of this work is to propose a novel method for training the ND using metric learning to improve performance and interpretability.
\end{enumerate}

At inference, the data are normalized before being passed through the AD. The anomalous data are then renormalized and pass through the ND, before being post-processsed and presented to operators for evaluation. See Figure \ref{fig:modelArch} for a visual depiction of this workflow. 

\subsection{Normalization and Training}
Training data for the AD consists of only benign network traffic which is normalized using a simple min-max scaler. Each dataset is scaled with independently to account for differences in baseline behavior patterns between networks. The detector then uses the Adam optimizer to minimize the mean-squared error reconstruction loss. 

In divergence from Redino et al. \cite{redino_2022}, the ND is trained on only known malicious behavior, independently normalized using a standard scaler followed by a Yeo-Johnson power transformation \cite{yeo_john}. The Novelty Detection loss function is a weighted sum of the reconstruction loss and a second loss function which implements metric learning on the latent vectors. Although many metric learning approaches to organize embeddings exist, we chose to use a triplet loss \cite{triplet_orig}, given by

\begin{equation}
\begin{split}
    \mathcal{L}(A, P, N) = \mathrm{max}(\lVert \mathrm{f}_{\phi}(A)-\mathrm{f}_\phi(P) \rVert ^{2}\\ - \lVert \mathrm{f}_{\phi}(A)-\mathrm{f}_{\phi}(N) \rVert ^{2} + \alpha, 0)  
\end{split}
\end{equation}

where $A$ is the anchor input, $P$ is the positive input, $N$ is the negative input, $\alpha$ is the margin, and $\mathrm{f}_{\phi}$ is an embedding function with trainable parameters. The goal for each triplet is to maximize the interclass distance, between anchor and negative, while minimizing the intraclass distance, between anchor and positive. A key step here is triplet mining, which creates batches of triplets according to a desired degree of hardness. Triplets with a larger distance between anchor and positive and lower distance between anchor and negative are considered harder, while the converse are considered easy. Convergence in training requires forming appropriately difficult batches so that the model can learn consistently and without too much instability. We used a 'round-robin' approach to create semi-hard batches, which satisfies equation \ref{eq:semi_hard} where $dist$ is the Euclidean distance, which contain approximately the same number of anchor examples for each pair of classes in order to encourage class separation even between classes with low support in training. Once a batch has been generated, the reconstruction loss is computed only on the anchor examples, and the triplet loss is calculated for the triplets after they have passed through the encoder.

\begin{equation} \label{eq:semi_hard}
    dist(A,P) < dist(A,N) < dist(A,P) + \alpha
\end{equation}

The final weighted loss equation for training taking into account the reconstruction loss $M$ and triplet loss $L$ is given by equation  \ref{eq:final_loss} where $\beta$ and $\gamma$ are scalar weights to bring each loss to approximately the same scale. Figure \ref{fig:tripletArch} depicts the final training architecture for the ND.

\begin{equation}
    \label{eq:final_loss}
    Loss = \beta\mathcal{M} + \gamma\mathcal{L} 
\end{equation}

\begin{figure*}[h!]
\caption{Novelty Detector Performance}
\label{table:overall_performance}
\centering
\begin{subfigure}{.5\textwidth}
    \caption{Reconstruction Loss Only}
    \centering
\begin{center}
\begin{tabular}{|c|c|c|c|c|} 
  \hline
  Attack & Support & AUC & Precision & Recall \\ 
  \hline
  rat & 940k & 0.84 & 0.69 & 0.45 \\
  infostealer & 430k & 0.94 & 0.56 & 0.29 \\
  command/control & 20k & 0.91 & 0.98 & 0.83 \\
  ransomware & 100k & 0.94 & 0.78 & 0.48 \\
  botnet & 100k & 0.95 & 1.00 & 0.91 \\
  interrogation & 130k & 0.91 & 0.91 & 0.92 \\
  worm & 100k & 0.97 & 0.83 & 0.53 \\
  downloader & 65k & 0.86 & 0.71 & 0.54 \\
  scanning & 170k & 0.86 & 0.94 & 0.91 \\
  \hline
  \textbf{Average} & - & \textbf{0.91} & \textbf{0.82} & \textbf{0.65} \\
  \hline
\end{tabular}
\end{center}
\label{fig:novelty_results}
\end{subfigure}%
\begin{subfigure}{.5\textwidth}
    \centering
\caption{Reconstruction Loss and Metric Learning}
\begin{center}
\begin{tabular}{|c||c|c|c|c|} 
  \hline
  Attack & Support & AUC & Precision & Recall \\ 
  \hline
  rat & 940k & 0.88 & 0.73 & 0.54 \\
  infostealer & 430k & 0.95 & 0.60 & 0.39 \\
  command/control & 20k & 0.91 & 0.99 & 0.84 \\
  ransomware & 100k & 0.87 & 0.80 & 0.56 \\
  botnet & 100k & 0.96 & 1.0 & 0.94 \\
  interrogation & 130k & 0.91 & 0.90 & 0.94 \\
  worm & 100k & 0.93 & 0.88 & 0.63 \\
  downloader & 65k & 0.86 & 0.77 & 0.61 \\
  scanning & 170k & 0.93 & 0.94 & 0.92 \\
  \hline
  \textbf{Average} & - & \textbf{0.91} & \textbf{0.85} & \textbf{0.71} \\
  \hline
\end{tabular}
\end{center}
\label{fig:e2e_results}
\end{subfigure}
\end{figure*}

\section{Experimental Design}

\subsection{Performance Metrics and Baseline Modeling}
Given the overwhelming bias toward benign events in real-world cybersecurity datasets, our primary performance metrics were precision, recall and the area under the curve (AUC) for the receiver operator characteristic (ROC) curve. We collaborated with the industry professional threat hunt teams to establish target values for these metrics that will allow the models to be useful in practice. To prevent operator fatigue, we aim first for high precision (0.90 or greater), and then attempt to maximize recall (0.80 or greater) and AUC. As compared to the performance of baseline modeling approaches such as support vector machines (SVMs) or multiclass random forests, the two-model architecture with graph features \cite{redino_2022} showed a significant improvement, with the AD performing with AUC above 0.99, especially when when evaluated on classes with low support or on novel network topologies. However, the performance of the ND lagged behind in these scenarios, struggling to identify novelty or conversely familiarity in attack types it had learned before if the number of training examples was low. Thus, the primary focus of these experiments was on the ND, though the AD is still necessary to generate end-to-end results.

\subsection{Anomaly Detector}
The AD was trained and evaluated using benign examples from OIF, MAWI, NCCDC, ISOT, and ISCX datasets and the training methodology aimed to minimize only a Mean Squared Error (MSE) loss without metric learning, primarily due to the fact that only a single class existed during training.

\subsection{Novelty Detector}
The overall classification power of the ND is measured by holding out a specific attack class during training, and then treating that class as a ZDT at prediction time. To quantify the change produced by metric learning, we treated each attack class separately as a ZDT and evaluated performance on the model trained with and without metric learning. Though not an explicit goal of this modeling effort, we are also interested in how well the encoder is able to separate different attack classes in the latent space. To that end, we used visualizations of the latent space and a $k$-nearest neighbors (kNN) classifier to identify whether one training methodology produced more distinct clusters than the other. 

\section{Implementation and Results}

\begin{figure*}[h!]
\centering
\begin{subfigure}{.5\textwidth}
    \centering
    \includegraphics[width=9cm, height=9cm]{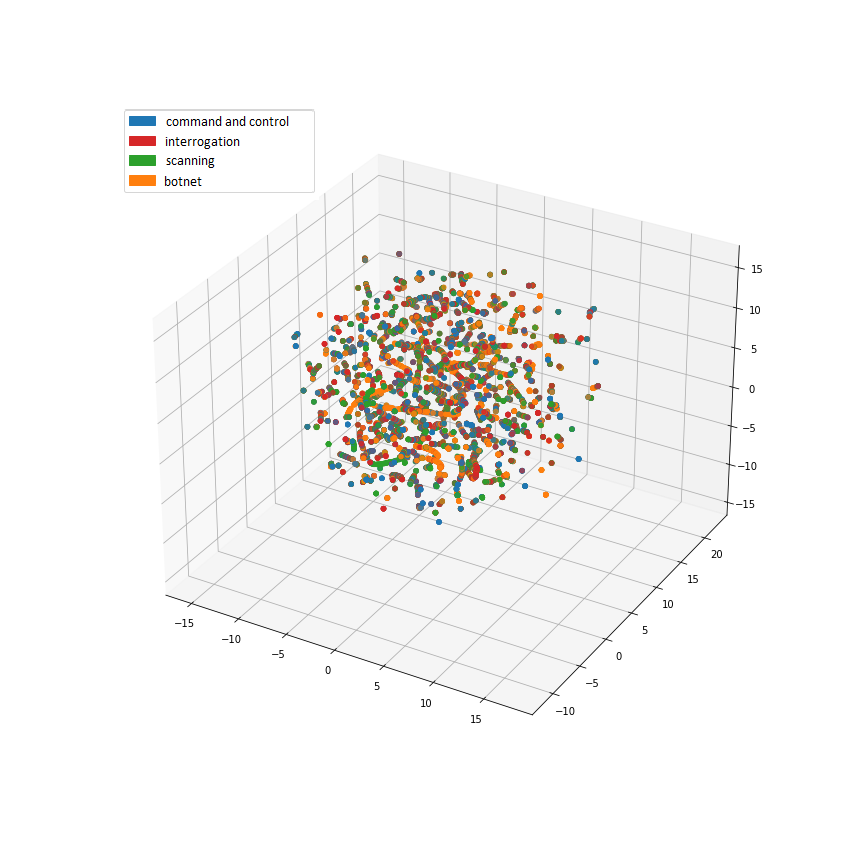}
\end{subfigure}\hspace{-1cm}
\begin{subfigure}{.5\textwidth}
    \centering
    \includegraphics[width=9cm, height=9cm]{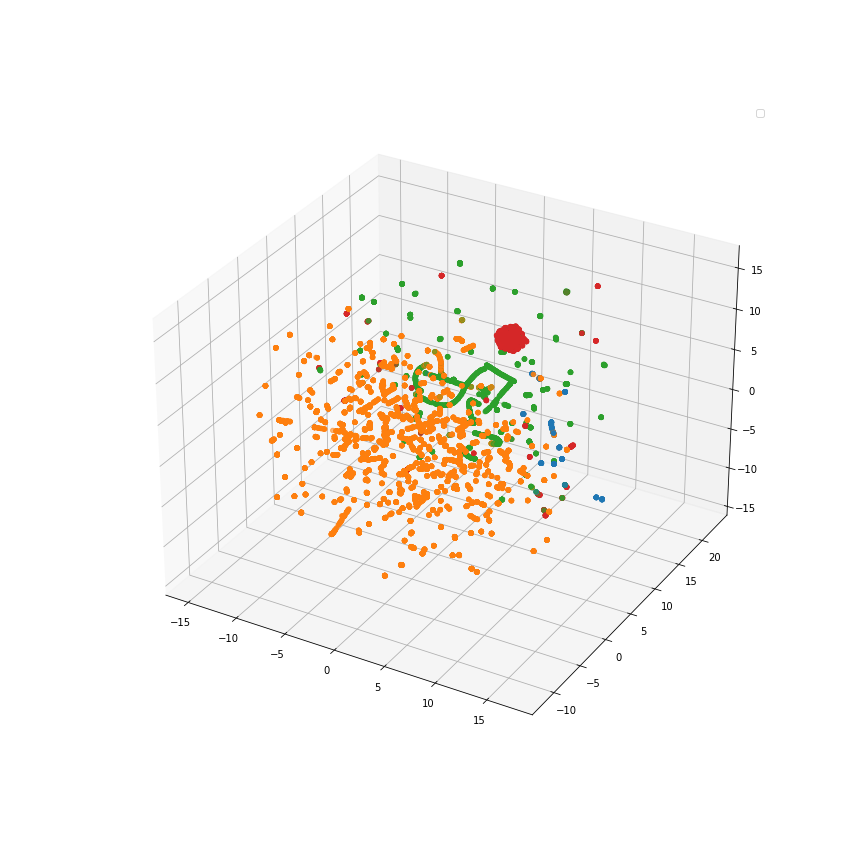}
\end{subfigure}
\vspace{-0.8cm}
\caption{Characteristic UMAP representations of the latent space for the model trained with only reconstruction loss (\textit{left}) and the model trained with metric learning (\textit{right}). Botnet attacks were treated as ZDTs and held out during training. Note that we only included the four attack classes found in NCCDC to allow for easier viewing.}
\label{fig:latent_space}
\end{figure*}

\subsection{Anomaly Detector}
Our results closely match Redino et al. \cite{redino_2022}, with precision, recall, and AUC above 0.99 for validation sets as well as completely held-out network datasets. 

\subsection{Novelty Detector}
The ND trained with the Codex, ISOT, and ISCX data, but without metric learning, also showed little change from \cite{redino_2022}. The version trained with metric learning, however, showed considerable improvement in precision and recall, particularly for classes with low support which had performed poorly in the previous model. Performance on each class as well as an overall average can be found in Table \ref{table:overall_performance}. In each case, in order to account for the rarity of ZDTs in real-world networks, the evaluation dataset contained less than $2\%$ events from the held-out class, and this class imbalance produces an apparent discrepancy between the measured AUC and precision-recall. We also held out the entire Codex dataset during training and were able to achieve similarly high performance at evaluation, demonstrating the model's ability to generalize between networks.

To visualize the latent spaces produced by these two training methodologies, we used a Uniform Manifold Approximation and Projection (UMAP) embedding to reduce the dimension of the space from 5 to 3  \cite{UMAP}. This necessarily destroys some information from the raw latent space, but it provides a more digestible visualization than simply projecting the space into every combination of 5 dimensions. A characteristic example of the difference between the two models, where the same attack class has been held out as a ZDT, can be seen in Figure \ref{fig:latent_space}. The model trained with metric learning shows significant organization of the known attack classes into clusters, while the held-out class, botnet, is more diffuse. The model trained with only reconstruction loss, by contrast, has a chaotic latent space with no clear clusters of any kind.

\begin{figure*}[h!]
\label{fig:latent_results}
\caption{Results for multiclass classification of known and holdout attacks using latent space representations and kNN.}
\centering
\begin{subfigure}{.44\textwidth}
    \centering
\includegraphics[width=8cm, height=5cm]{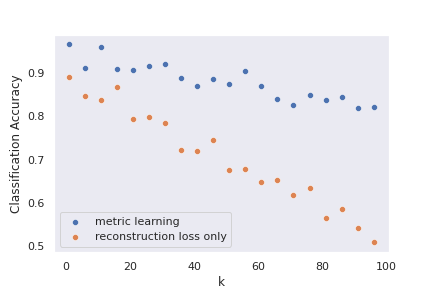}
\caption{kNN classification performance for both versions of the ND at different values of $k$.}
\label{fig:knn}
\end{subfigure}%
\hspace{.05\textwidth}
\begin{subfigure}{.44\textwidth}
    \centering
\begin{center}
\begin{tabular}{|c|c|c|}
\hline
\textbf{Holdout} & \textbf{CAT} & \textbf{Probability} \\
\hline
scanning & interrogation & 0.34 \\
botnet & command/control & 0.32 \\
command/control & downloader & 0.29 \\
infostealer & downloader & 0.28 \\
\hline
\end{tabular}
\end{center}
\caption{Closest attack type attribution for various holdout examples. The probability is based on the number of nearest neighbors for the chosen attack class compared to all others.}
\label{fig:cata}
\end{subfigure}
\end{figure*}

In order to test whether this clustering effect occurred in the raw, 5-dimensional latent space, we trained the model with all available attack classes, then fit a kNN to the embeddings and attempted to classify test examples of the known attack classes. Figure \ref{fig:knn} shows the classification accuracies for both versions of the model over a range of $k$ values. For small $k$, both models performed well, with average accuracies of $0.88$ and $0.93$ for the models trained without and with metric learning, respectively. As the value of $k$ increased, though, the performance of the model trained with only reconstruction loss degraded much more rapidly than the metric loss model, demonstrating the more robust clustering produced by the triplet loss.
    
\subsection{End-to-End Performance}
Because of the high performance of the AD, the overall performance of the dual-autoencoder approach is almost identical to that of the ND. For the version of the model where the ND was trained without metric learning, the precision and recall were $0.82$ and $0.64$, respectively. Meanwhile, for the version where the ND is trained with metric learning, the precision and recall were $0.85$ and $0.71$, respectively, showing a clear improvement over the previous version. While the improvement in performance may be modest compared to the datasets used for evaluation here, we believe that this will provide a drastic improvement in end-user usability in enterprise environments where the number of events processed by the models would be several orders of magnitude larger. Furthermore, given the rapidly increasing number of ZDTs seen in industry year-over-year \cite{dunsavage_2022}, a \textit{7\%} improvement in recall could enable organizations to identify potentially severe ZDTs that would have never been flagged by previous methods.

\section{Closest Attack Type Attribution (CATA)}
While utilizing metric learning lends itself to performance improvements when identifying ZDTs, the class separation in the ND's latent space also allows us to improve model interpretability from an end-user perspective. Particularly, we can utilize a k-nearest neighbors classification in the ND's latent space to identify the closest attack type to a ZDT example during inference using the algorithm given below.

\begin{algorithm}
\begin{algorithmic}[1]
\caption{Closest Attack Type Attribution}
\Require $K$, the kNN model fit to the ND while training
\For{Example $e$ in batch}
\State $c$ = latent space representation for $e$ \Comment{Pass example through encoder}
\State $p$ = $K$($c$) \Comment{Compute closest attack type for $e$ using kNN with probability}
\EndFor
\end{algorithmic}
\end{algorithm}

The output of the CATA algorithm would be a per-example attribution of the  closest attack class the ND had learnt to encode during training as seen in Figure \ref{fig:cata} for some holdout examples. We believe that generating such classification probabilities of related attack types provides deeper insight into model interpretability and evaluation in two distinct ways. Firstly, we can evaluate the stability and efficacy of model training by measuring the accuracy of kNN classification with varying values of $k$. Secondly, we can test the ND's \lq conceptual \rq{} understanding of cyber attack types by holding out certain classes during training and evaluating the CATA output for those holdouts to measure the classification probability for training classes relative to it. Ideally, learned classes that are similar to the holdout will have higher CATA probabilities as compared to dissimilar classes. Naturally, the fidelity of the CATA algorithm would be determined by the diversity of classes seen during training time but the strong kNN classification accuracy observed during experimentation even with large $k$ provides a promising indication of reliably stable results during inference. 

Furthermore, from a threat hunt perspective, this metadata should allow for faster investigation of novel threats by providing crucial context related to expected behavior of malware, possibly leading to fruitful avenues of exploration; ultimately to help reduce the mean time to triage and respond in SOC organizations. CATA probabilities could also help SOC operators quickly triage false positives our models produce in production by validating if the flagged ZDT is, in reality, the closest CATA attack.

\section{Conclusion, Limitations, and Future Work}
The introduction of metric learning to the ND presented here offers a modest but noticeable improvement over the previous version trained with only reconstruction loss. It is important to note here that while triplet loss based metric learning only provides a few percentage points of improvement in precision and recall, the framework introduced in this work can be used to implement other metric learning methods as well. Our team has also specifically experimented with softmax-based metric learning methods and have observed \textbf{precision values ranging from 0.72 to 0.97 and recall from 0.58 to 0.97}, which is a drastic improvement to the triplet loss method.

While the increase in overall precision and recall is important to avoid operator fatigue, perhaps equally valuable is the ability to better interpret individual model outputs. Threat hunters tasked with validating ZDT detections from our model are able to see whether the event was similar to a known attack class based on the output of kNN in the latent space; more nearest neighbors from a particular class corresponds to a higher \lq\lq probability\rq\rq{} that the supposed zero-day event was actually an attack of that known class. This offers threat hunters some indication of why the model made its prediction, and by giving a starting point for investigation it has the potential to significantly reduce the time required to make a decision.

While we made every effort to use a variety of attack types, the field of known attacks is more diverse than the labeled data available to us. Thus, in practice many attacks would be falsely labeled as ZDTs simply because they were unknown to the network in training. We expect the frequency of false positives to decrease as operators label the output data and it is used to retrain the model with new attack types. Furthermore, we believe that continued research in cyber terrain mapping and annotation could greatly alleviate this problem by providing our models contextual labels of cyber threats based on the risk an individual asset in the flow poses to the network as a whole.

Another potential issue here is a specific form of overfitting where the latent space collapses into a smaller dimension, destroying necessary information for the decoder. This happened occasionally during training with the triplet loss and it was vital to check for it. Other loss functions are available for implementing metric learning, such as the softmax-based function mentioned above, and have so far avoided this overfitting problem.  Future work will include continued testing of other loss functions, development of federated learning, and development of metric based few shot learning techniques that allow our models to learn across environments with limited labelled attack examples while maintaining data privacy. 

Future work will also aim to benchmark our models' performance compared to previously introduced approaches on the same datasets we used for training in order to achieve a more in-depth comparison of results. Particularly, we believe that the works introduced by Hindy et al.\cite{vanillaAutoEnc}, Yousefi-azar et al.\cite{yousefiAE}, and An and Cho\cite{VAE} would be feasible to implement and benchmark against given the data sources the authors used in their studies as well as similar themes in terms of a modeling approach.

\bibliographystyle{IEEEtran}
\bibliography{ref}

\end{document}